\documentclass[twocolumn,amsmath,amssymb]{revtex4}

\usepackage{graphicx}

\begin{document}

\title{Optical solitons in graded-index multimode fiber}

\author{W. H. Renninger}
\email{whr6@cornell.edu}

\author{F. W. Wise}
\email{frank.wise@cornell.edu}

\affiliation{School of Applied and Engineering Physics, Cornell University, Ithaca, New York 14853}

\begin{abstract}
Solitons are non-dispersing localized waves that occur in diverse physical settings. A variety of optical solitons have been observed, but versions that involve both spatial and temporal degrees of freedom are rare.  Optical fibers designed to support multiple transverse modes offer opportunities to study wave propagation in a setting that is intermediate between single-mode fiber and free-space propagation. Here we report the observation of optical solitons and soliton self-frequency shifting in graded-index multimode fiber.  These wave packets can be modeled as multi-component solitons, or as solitons of the Gross-Pitaevskii equation. Solitons in graded-index fibers should enable increased data rates in low-cost telecommunications systems, are pertinent to space-division multiplexing, and can offer a new route to mode-area scaling for high-power lasers and transmission.
\end{abstract}

\maketitle

\titlepage

\twocolumngrid

Solitons are localized waves that arise from the interplay of linear and nonlinear processes that individually would cause the wave to decay. They occur in numerous physical settings, including liquids \cite{Russel1844,Osborne1980}, optical fibers \cite{moll}, plasmas \cite{Zabusky1965},  and condensed matter \cite{Khaykovich2002, Demokritov2003}. Temporal solitons that form in single-mode optical fiber \cite{Hasegawa, moll} are perhaps the quintessential example, and they have had major influence on telecommunications.  The potential impact of soliton formation in multimode fiber was appreciated by early workers \cite{Crosignani:81,Hasegawa:80,Crosignani:82}.  More recently, multimode fiber has been considered theoretically as an environment that could support spatiotemporal solitons  \cite{Yu1995167,Raghavan2000377}, or light bullets, which attract interest owing to their particle-like nature and potential for all-optical switching \cite{sdragginglogic,Liu1999, PhysRevLett.105.263901}. Perhaps surprisingly, no experiments have been reported despite the theoretical progress.

Optical pulse propagation in a multimode fiber involves an intricate mix of spatiotemporal phenomena coupled fundamentally by nonlinearity and practically by waveguide imperfections. Modeling of the pulse propagation \cite{Crosignani:81,Hasegawa:80,Crosignani:82,Poletti:08,6242367} is difficult, and multiple approaches have been taken. These include rewriting the coupled modes in terms of principal modes \cite{Shemirani2009,Shemirani2009a,Shen:05}, variational solution of a nonlinear wave equation \cite{Yu1995167,Raghavan2000377,Longhi2004}, and analysis of optical-wave thermalization and condensation \cite{PhysRevA.83.033838}.  From a modal perspective, soliton formation requires nonlinear coupling between the modes to cancel the effects of modal dispersion \cite{Crosignani:81,Hasegawa:80,Crosignani:82}. Such multi-component or vector solitons have been studied in several contexts \cite{Barad1997,Kang1996,Chen1997}. There is only one prior report of spatiotemporal vector solitons \cite{Liu1999}, where the two components were different colors. Alternatively, nonlinear pulse pulse propagation in a multimode waveguide can be analyzed with a three-dimensional Gross-Pitaevskii equation \cite{Pitaevskii2003}, which is widely used to model Bose-Einstein condensates.  The soliton solutions simultaneously balance nonlinearity with diffraction, dispersion, and a spatial harmonic potential.

In addition to its intrinsic scientific interest, the formation of solitons in graded-index (GRIN) fiber will be relevant to applications. Owing to their low cost and ease of alignment, multimode fibers are widely used in high-speed local area networks \cite{Agrawal1997}.  The maximum data rate is limited by intersymbol interference that arises from modal dispersion.  A new pulse propagation technique that can retain the simplicity of multimode systems while avoiding modal dispersion should be beneficial to low-cost, high-speed systems.  As systems approach the Shannon limit for information transmission, interest in exploiting multiple spatial channels \cite{Berdague1982,Stuart2000,Tarighat2007}, which could be transverse modes, has grown.  Whether used to minimize cross-talk in neighboring modes or to determine nonlinear limits to such approaches, soliton formation will be a factor in the design of such technologies.  Finally, there is a growing need for large-mode-area fibers for the generation and transmission of pulses with ever-higher peak powers \cite{Fermann1998}.  Soliton transmission through large-core graded-index fibers could play a major role in these applications.

In this Article, we describe theoretical and experimental observations of optical solitons and soliton self-frequency shifting in GRIN multimode fiber. Remarkably, the stable solutions of the coupled-mode equations and the Gross-Pitaevskii equation are equivalent.  Implications of the results for telecommunication, space-division multiplexing, and high-power laser and transmission systems are discussed.

\begin{figure}[htb]
\centerline{\includegraphics[width=8.0cm]{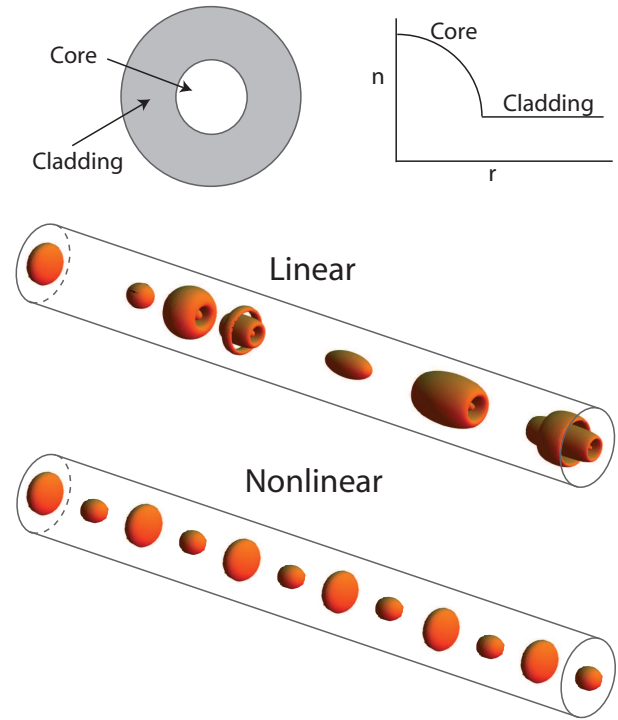}}
\caption{\textbf{Soliton formation in GRIN fiber:} Top: Schematic of the core and cladding of the fiber and the relative index as a function of radius.  Middle: In linear propagation, the spatial modes separate in time, and individually increase in duration. Bottom: Propagation of the soliton.}
\label{Conceptfig}
\end{figure}

\section*{\textbf{Results}}
\subsection*{\textbf{Coupled-mode analysis}}
The most-complete theoretical model is based on coupled modes of the electromagnetic field \cite{Poletti:08,6242367}. The complex electric field can be decomposed into a sum of spatial functions for the modes, $F_p(\rho,\phi,\omega)$, multiplied by the evolving envelopes, $A_p(z,\omega)$:

\begin{equation}
\begin{aligned}
E(\rho,\phi,\omega)=\sum_p F_p(\rho,\phi,\omega)e^{i\beta_p(\omega)z}A_p(z,\omega).
\end{aligned}
\label{fullfieldeq}
\end{equation}

\begin{figure*}[htb]
\centerline{
\includegraphics[width=14.0cm]{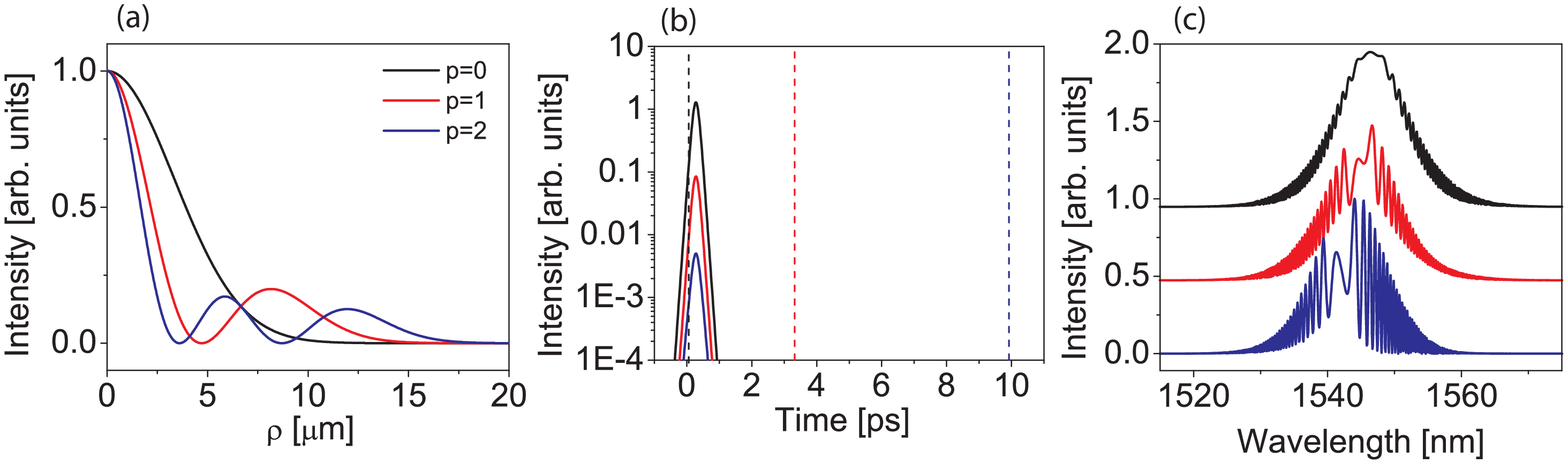}}
\caption{\textbf{Multi-component soliton:} (a) The three spatial modes and the corresponding (b) temporal pulse profiles and (c) \textbf{normalized} spectra.  The dotted lines correspond to the location of the pulse center after linear pulse propagation (without nonlinear attraction between modes).}
\label{vectorfig}
\end{figure*}

To simplify the problem, we consider only the radially-symmetric modes (written explicitly in the Supplementary Information), which are relevant to our experiments. The normalized modes, $F_p(\rho,\omega)$, and corresponding propagation constants, $\beta_p(\omega)$, can be written as \cite{6242367}
\begin{equation}
\begin{aligned}
&F_p(\rho,\omega)=\sqrt{\frac{2}{\pi \text{w}_0^2}}e^{-\frac{\rho^2}{\text{w}_0^2}}L_p\left(2\frac{\rho^2}{\text{w}_0}\right),\text{ and}\\
&\beta_p(\omega)=k\sqrt{1-\frac{2}{k}\frac{\sqrt{2\Delta}}{R}(2p+1)},
\end{aligned}
\end{equation}
where $L_p$ is a Laguerre polynomial, $k=\omega n_0/c$,  $\Delta$ is the index difference between the center and the cladding of the fiber, $R$ is the fiber core radius, and $\text{w}_0=(2R^2/k^2\Delta)^{1/4}$ is the fundamental mode size.  By neglecting higher-order dispersive and nonlinear effects, the equations can then be written as
\begin{equation}
\begin{aligned}
\frac{\partial A_p(z,t)}{\partial z}=&i\delta\beta_0^{(p)}A_p-\delta\beta_1^{(p)}\frac{\partial A_p}{\partial t}-i\frac{\beta_2}{2}\frac{\partial^2 A_p}{\partial t^2}\\&+i\frac{\gamma}{\pi \text{w}_0^2}\sum_{l,m,n}\eta_{plmn}A_lA_mA^*_n,
\end{aligned}
\label{coupledeq}
\end{equation}
where $\delta\beta_0^{(p)}$ ($\delta\beta_1^{(p)}$) is the difference between the $p_{th}$ and the fundamental mode of the first (second) coefficient in the Taylor expansion of $\beta_p$ about $\omega_0$.  $\beta_2$ corresponds to the material group-velocity dispersion, $\gamma=\omega_0 n_2/c$ is the nonlinear coefficient, and $\eta_{plmn}$ are the nonlinear coupling coefficients (defined explicitly in the Supplementary Information). The equations are solved numerically for a variety of input fields launched into 100 m of a standard GRIN fiber (parameters in Methods), to correspond to experiments described below. The number of modes, and hence equations, required to account for the pulse propagation depends on the modes that are initially seeded. For the cases studied here the GRIN fiber is seeded by the output of a standard single-mode fiber and $>99.9\%$ of the energy is accounted for with only the 3 lowest order symmetric modes (Figure \ref{vectorfig}(a)).

\begin{figure*}[htb]
\centerline{
\includegraphics[width=18.0cm]{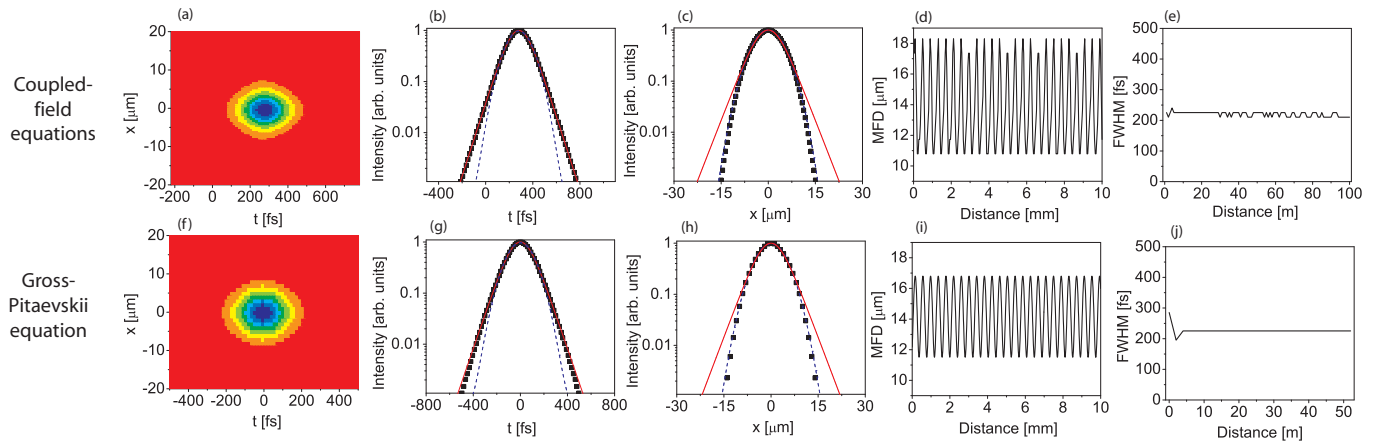}}
\caption{\textbf{Simulation results:}  Numerical solution of the coupled-field equations (a-e) and the Gross-Pitaevskii equation (f-j). (a) Space-time intensity plot, (b) pulse averaged over space and (c) beam averaged over time (symbols) with a hyperbolic secant squared (solid line) and Gaussian (dashed line) fits,  evolution of the (d) spatial averaged over the pulse and of the (e) FWHM pulse duration averaged over the beam from coupled-field equations and (f)-(j) the corresponding plots for the Gross-Pitaevskii equation.}
\label{bigfig}
\end{figure*}

In linear propagation, the modes separate in time owing to their different group velocities and the pulse duration in each mode increases because of material dispersion (middle panel of Figure 1). The beam waist oscillates with period $\frac{\pi R}{\sqrt{2\Delta}}$ until the pulses separate temporally, after which the oscillation ceases and the spatial evolution is essentially that of the fundamental mode. At higher powers, nonlinearity balances the material dispersion, mode-coupling counteracts the group-velocity mismatch, and a multi-component soliton forms. In other words, the duration and temporal location of light energy in each mode are locked, and remain unchanged with propagation. While the existence of solitons in multimode systems was suggested based on analytical arguments \cite{Crosignani:81,Hasegawa:80,Crosignani:82}, this is the first demonstration that multi-component solitons are even theoretically stable in a complete numerical model of the coupled modes. The result of launching a 300-fs, $\sim$0.5-nJ pulse is shown in Figure \ref{vectorfig}(b), e.g. The three modes overlap in time, with the group delay equal to an energy-weighted average of the modal delays. The spectra of the individual modes shift to have the same  group velocity, with the higher-order modes blue-shifted (Figure \ref{vectorfig}(c)). The structured spectra result from radiation of energy from each mode as the soliton forms. Similar structure appears when solitons form in single-mode fiber (Supplementary Figure 2 in the Supplementary Information.) The space-time profile is nearly symmetric (Figure \ref{bigfig}(a)), with a sech$^2$ temporal intensity profile (Figure \ref{bigfig}(b)) and a Gaussian spatial profile (Figure \ref{bigfig}(c)).  The mode-field diameter (MFD) averaged over the pulse (Figure \ref{bigfig}(d)) oscillates around the fundamental mode size ($2\text{w}_0$) with a period equal to that of low-intensity light. The full-width at half-maximum (FWHM) pulse duration averaged over the beam (Figure \ref{bigfig}(e)) quickly converges to a steady solution that is $\sim10$ times shorter than the output pulse duration would be due to group-velocity dispersion alone.

\subsection*{\textbf{Single-field analysis}}
Alternatively, the system can be analyzed with a nonlinear wave equation for the total field. Standard procedures are used to reduce the Maxwell equations to a single wave equation.  By employing the paraxial and slowly-varying envelope approximations, a Gross-Pitaevskii equation is obtained \cite{Raghavan2000377,Yu1995167}:
\begin{equation}
\begin{aligned}
\frac{\partial A}{\partial z}=&\frac{i}{2k_0}\left(\frac{\partial^2 A}{\partial x^2}+\frac{\partial^2 A}{\partial y^2} \right)-i\frac{\beta_2}{2}\frac{\partial^2 A}{\partial t^2}\\
-&i\frac{k_0\Delta}{R^2}(x^2+y^2)A+i\gamma|A|^2A.
\end{aligned}
\label{3deq}
\end{equation}
where $A$ is the slowly-varying envelope at center frequency $\omega_0$ and t is time in a reference frame moving at the group velocity of the pulse. The paraxial approximation does neglect the variation of group velocities among the modes. The approximation is justified in light of the solution to the coupled-field equations, which shows the modes aligning to a common group velocity.  To reach manageable computational times, we reduce the system to the 2-dimensional case with y=0 as in Ref. \cite{Raghavan2000377}.  The solution quickly converges and simulations are stopped after 52 m of propagation (which requires 20 days of computation).  Remarkably, the soliton solution is nearly identical to that found with the coupled-mode equations (Figure \ref{bigfig}), except for a slight decrease in the amplitude of the spatial oscillation (Figure \ref{bigfig}(i)).

The identification of terms in Equation \ref{3deq} with specific physical processes provides insight, but exact analytic solutions are not known. An approximate analytic solution is obtained by making the variational approximation with the trial function
\begin{equation}
\begin{aligned}
A(x,y,z,t)&=\sqrt{\frac{E}{2\pi \text{w}_0^2 \tau}}Sech\left(\frac{t}{\tau}\right)\\
&\times e^{\frac{-(x^2+y^2)}{\text{w}_0^2}+i(\theta t^2+\alpha(x^2+y^2)+\phi)},
\end{aligned}
\label{varansatz}
\end{equation}
where $E=\int|A|^2dxdydt$ is the energy, $\tau$ is the pulse duration, $\text{w}_0$ is the beam width, $\theta$ and $\alpha$ are chirp parameters, and $\phi$ is an arbitrary phase  \cite{Kivshar2003,Raghavan2000377,Yu1995167}. If the quantity $\epsilon=\frac{E^2\gamma^2k_0^2\sqrt{2\Delta}}{24\pi^2|\beta_2|R}$ is much less than one, a stable fixed point to the equations of motion is given by
\begin{equation}
\text{w}_0=(2R^2/k_0^2\Delta)^{1/4}\;\text{and}\;E\tau=\frac{2|\beta_2|c \pi\text{w}_0^2}{\omega_0 n_2}.
\label{areathm}
\end{equation}
For typical fiber parameters $\epsilon=2\times10^{-4}$, so the approximation is excellent.  The fixed point is thus the fundamental mode of the GRIN fiber with a temporal profile that corresponds to the soliton with that beam size.

\subsection*{\textbf{Experiments}}
\begin{figure}[htb]
\centerline{
\includegraphics[width=8.0cm]{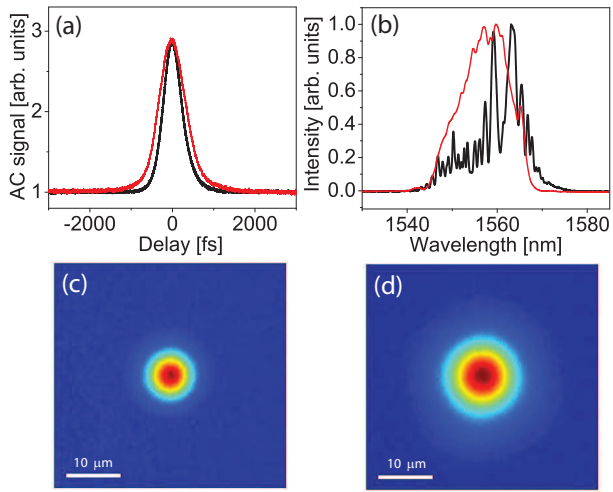}}
\caption{\textbf{Soliton formation with 0.5-nJ pulse energy:} (a) Autocorrelation trace of the pulse and (b) spectrum before (red) and after (black) the GRIN fiber and beam profile (c) before and (d) after the GRIN fiber.}
\label{500pJex}
\end{figure}

\begin{figure}[htb]
\centerline{
\includegraphics[width=8.0cm]{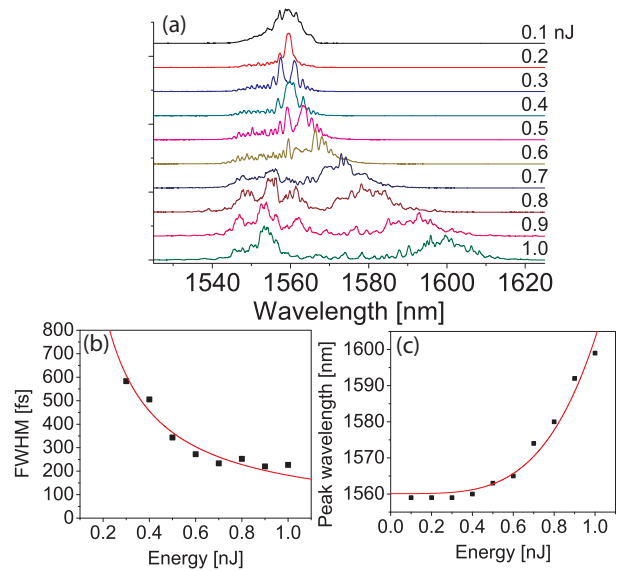}}
\caption{\textbf{Experimental trends with pulse energy:} (a) Output spectra for indicated pulses energies.  (b) Measured pulse duration vs. energy (symbols) and fit to Equation \ref{areathm}.  (c) Measured peak wavelength vs. pulse energy (symbols) and fit to Equation \ref{sfseq}.}
\label{results}
\end{figure}

The theoretical results suggest that the required pulse energy at 1550 nm can be reached with readily-designed Er-doped fiber lasers.  Experiments were performed with a source that generates $\sim300$-fs pulses (red lines in Figure \ref{500pJex}(a) and (b)) with energy up to $\sim3$ nJ. The initial spatial profile is the fundamental mode of an ordinary single-mode fiber with MFD of 11.5 $\mu$m (Figure \ref{500pJex}(c)). With these parameters, 100 m of GRIN fiber comprises $\sim100$ dispersion lengths (i.e., in linear propagation the pulse will broaden about 100 times), $\sim400$ nonlinear lengths, and $\sim1,000,000$ diffraction lengths. For energies below 0.3 nJ the pulse disperses, and it is difficult to measure the autocorrelation of the output pulse. Results of launching a 0.5-nJ pulse illustrate soliton formation (Figure \ref{500pJex}).  After propagation through the fiber, the pulse is compressed temporally (Figure \ref{500pJex}(a)) and the spectrum becomes structured (Figure \ref{500pJex}(b)). The MFD at the end of the GRIN fiber is 17.9 $\mu$m (Figure \ref{500pJex}(d)). At higher energies, the output pulse duration decreases, and the spectrum broadens and red-shifts (Figure \ref{results}(a)), in a manner immediately reminiscent of soliton self-frequency shifting \cite{Mitschke1986}.

We can conveniently analyze the experimental results using the results of the variational approach. The dependence of pulse duration on pulse energy is predicted well by the fixed point (Figure \ref{results}(b)): the best fit is obtained with a MFD of 17.4 $\mu$m, which is close to the measured value of 17.9 $\mu$m. For the spectral shift we apply the results of standard soliton perturbation theory. The wavelength shift is inversely proportional to the fourth power of the  pulse duration \cite{Mitschke1986}:
\begin{equation}
\Delta\lambda=\frac{4\lambda^2 \beta_2 T_R z}{15 \pi c \tau^4}.
\label{sfseq}
\end{equation}
Here $z$ is distance and $T_R$ is related to the slope of the Raman gain spectrum.  The pulse duration is inversely proportional to the energy, so the wavelength shift should be proportional to the fourth power of the energy. Indeed this is the case (Figure \ref{results}(c)): the best fit is obtained with $T_R=2.6$ fs, which is close to the accepted value of 3 fs for silica fiber.

\section*{\textbf{Discussion}}
The spatial and temporal degrees of freedom are coupled through the Kerr nonlinearity in a multimode waveguide. The picture that emerges from the theoretical and experimental results presented above is that a 3-component soliton forms with nonlinear pulse propagation. That is, the process that is responsible for compensation of group-velocity dispersion in time also compensates for the separation of the different spatial modes (modal dispersion).
The spatial evolution of the solitons in GRIN fiber has a significant linear contribution, as expected with a waveguide.  However, the system differs significantly from a single-mode waveguide because the spatial dimensions have the freedom to play a major role.  Therefore, it is surprising that, on average (ignoring spatial oscillations), the resulting solutions are equivalent to single-mode solitons propagating in the fundamental mode of the fiber. This is noteworthy considering that the system includes multiple modes (Equation \ref{coupledeq}), and is modeled by a three-dimensional Gross-Pitaevskii equation (Equation \ref{3deq}).

Experimentally, measurements of the beam propagation on a sub-millimeter length scale will be valuable to confirm the presence of multiple transverse modes, but we are unable to make such measurements with adequate accuracy.  While the analyses presented above clearly account well for the experiments, we cannot absolutely rule out an alternative explanation. Although the models would become quite complex, Raman scattering could be incorporated directly, rather than treated as a perturbation. In such a model, the interplay of linear mode coupling (which distributes energy uniformly among modes) and Raman scattering could result in energy being ``siphoned'' into the fundamental mode of the fiber \cite{Bespalov1978,Goldhar1982,Chang1983}. The occurrence of this process would be noteworthy in its own right.  We hope that the results presented here stimulate efforts to address this intriguing spatiotemporal problem.

Regardless of the final theoretical interpretation, the observation of optical solitons and soliton self-frequency shifting in a GRIN waveguide has implications for a variety of scientific and technological issues.  GRIN media offer a stable and convenient setting for the study of spatiotemporal wave propagation, with solutions that also pertain to Bose-Einstein condensates.  Soliton transmission in multimode fibers relaxes the stringent coupling requirements for single-mode systems, and thereby could reduce the cost of high-data-rate telecommunication systems \cite{Agrawal1997}.  GRIN solitons will be pertinent to current efforts to reach the Shannon limit \cite{Essiambre2010} through space-division multiplexing \cite{Berdague1982,Stuart2000,Tarighat2007}.  In such systems, cross-talk can be minimized by utilizing modes that do not overlap spatially as channels. Finally, the linear scaling of soliton energy and power with core diameter will benefit applications in high-power lasers and pulse transmission \cite{Fermann1998}.  For example, commercially-available GRIN fiber with 1-mm core diameter will allow transmission of 100-kW solitons.

\section*{\textbf{Methods}}


For the coupled equation simulations, $\lambda_0=1550$ nm, E $=0.5$ nJ, $n_2=3.2\times10^{-20}$ m$^2$/W, $n_0=1.444$, $\beta_2=-281$ fs$^2$/cm, $\Delta=0.0029$, and $R=31.25$ $\mu$m.  An overlap-integral calculation with a 10-$\mu$m  Gaussian input reveals that $>99.9\%$ of the energy is accounted for with 3 modes ($92.22\%$ in $p=0$, $7.17\%$ in $p=1$, and $0.56\%$ in $p=2$).  In linear propagation, the pulse in the $p=1$ mode moves away from the pulse in the fundamental mode at a rate of 33 fs/m (99 fs/m for $p=2$).

We use a split-step Fourier-transform method to numerically solve Equation \ref{3deq}.  With 1024 points in time and 512 in space, and a 100-$\mu$m step in the propagation direction, simulation of pulse propagation through 52 m of fiber requires 20 days of computation using four cores on an Intel i7 computer. The parameters used for the simulations are the same as those for the coupled system, with an effective length of 11.8 $\mu$m for the y dimension.


Experiments were performed at 1550-nm wavelength where the group-velocity dispersion of fused silica is anomalous.  An Erbium-doped fiber oscillator (11 MHz repetition rate) operating in the normal-dispersion regime produces $\sim300$-fs pulses with $\sim1$-nJ pulse energy.  The oscillator is followed by 80 m of normal-dispersion fiber stretcher, a single-mode amplifier, and a transmission grating compressor in a standard chirped-pulse amplifier configuration.  The output is aligned into a SMF-28 fiber pigtailed collimator of 50-cm length, which allows for near ideal seeding conditions when it is spliced directly to the GRIN fiber (Thorlabs GIF625).  After all losses due to the grating compressor and coupling into the collimator, this setup allows for up to $\sim3$ nJ at a dechirped duration of $\sim300$ fs.

The pulse is directly measured with a 2-photon intensity autocorrelator, the spectrum is measured with a grating spectrometer with $0.4$-nm resolution, and the beam is measured with an InGaAs camera with a 111x beam expander to fill the detector array.  The MFD is measured by averaging the widths of a Gaussian fit of the x and y cross-sections at the center of the beam.  The pulse duration as a function of energy for the pulses after the GRIN fiber are measured by taking the full-width at half-maximum duration of an intensity autocorrelation and dividing by the correlation factor for a sech$^2$ intensity profile (1.543).


\begin{thebibliography}{10}

\bibitem{Russel1844}
John~Scott Russell.
\newblock {Reports on waves}.
\newblock {\em Rep. Br. Assoc. Adv. Sci.}, pages 311--390, 1844.

\bibitem{Osborne1980}
A~R Osborne and T~L Burch.
\newblock {Internal solitons in the andaman sea.}
\newblock {\em Science (New York, N.Y.)}, 208(4443):451--60, May 1980.

\bibitem{moll}
L~F Mollenauer, R~H Stolen, and J~P Gordon.
\newblock {Experimental Observation of Picosecond Pulse Narrowing and Solitons
  in Optical Fibers}.
\newblock {\em Phys. Rev. Lett.}, 45(13):1095--1098, September 1980.

\bibitem{Zabusky1965}
N.~Zabusky and M.~Kruskal.
\newblock {Interaction of "Solitons" in a Collisionless Plasma and the
  Recurrence of Initial States}.
\newblock {\em Physical Review Letters}, 15(6):240--243, August 1965.

\bibitem{Khaykovich2002}
L~Khaykovich, F~Schreck, G~Ferrari, T~Bourdel, J~Cubizolles, L~D Carr,
  Y~Castin, and C~Salomon.
\newblock {Formation of a matter-wave bright soliton.}
\newblock {\em Science (New York, N.Y.)}, 296(5571):1290--3, May 2002.

\bibitem{Demokritov2003}
Sergej~O Demokritov, Alexander~A Serga, Vladislav~E Demidov, Burkard
  Hillebrands, Michail~P Kostylev, and Boris~A Kalinikos.
\newblock {Experimental observation of symmetry-breaking nonlinear modes in an
  active ring.}
\newblock {\em Nature}, 426(6963):159--62, November 2003.

\bibitem{Hasegawa}
A~Hasegawa and F~Tappert.
\newblock {Transmission of stationary nonlinear optical pulses in dispersive
  dielectric fibers. I. Anomalous dispersion}.
\newblock {\em Appl. Phys. Lett.}, 23:142, 1973.

\bibitem{Crosignani:81}
Bruno Crosignani and Paolo~Di Porto.
\newblock {Soliton propagation in multimode optical fibers}.
\newblock {\em Opt. Lett.}, 6(7):329--330, July 1981.

\bibitem{Hasegawa:80}
Akira Hasegawa.
\newblock {Self-confinement of multimode optical pulse in a glass fiber}.
\newblock {\em Opt. Lett.}, 5(10):416--417, October 1980.

\bibitem{Crosignani:82}
Bruno Crosignani, Antonello Cutolo, and Paolo~Di Porto.
\newblock {Coupled-mode theory of nonlinear propagation in multimode and
  single-mode fibers: envelope solitons and self-confinement}.
\newblock {\em J. Opt. Soc. Am.}, 72(9):1136--1141, September 1982.

\bibitem{Yu1995167}
Shinn-Sheng Yu, Chih-Hung Chien, Yinchieh Lai, and Jyhpyng Wang.
\newblock {Spatio-temporal solitary pulses in graded-index materials with Kerr
  nonlinearity}.
\newblock {\em Optics Communications}, 119(1–2):167--170, 1995.

\bibitem{Raghavan2000377}
S~Raghavan and Govind~P Agrawal.
\newblock {Spatiotemporal solitons in inhomogeneous nonlinear media}.
\newblock {\em Optics Communications}, 180(4–6):377--382, 2000.

\bibitem{sdragginglogic}
Robert McLeod, Kelvin Wagner, and Steve Blair.
\newblock (3+1)-dimensional optical soliton dragging logic.
\newblock {\em Phys. Rev. A}, 52(4):3254--3278, October 1995.

\bibitem{Liu1999}
X.~Liu, L.~Qian, and F.~Wise.
\newblock {Generation of Optical Spatiotemporal Solitons}.
\newblock {\em Physical Review Letters}, 82(23):4631--4634, June 1999.

\bibitem{PhysRevLett.105.263901}
S~Minardi, F~Eilenberger, Y~V Kartashov, A~Szameit, U~R\"{o}pke, J~Kobelke,
  K~Schuster, H~Bartelt, S~Nolte, L~Torner, F~Lederer, A~T\"{u}nnermann, and
  T~Pertsch.
\newblock {Three-Dimensional Light Bullets in Arrays of Waveguides}.
\newblock {\em Phys. Rev. Lett.}, 105(26):263901, December 2010.

\bibitem{Poletti:08}
Francesco Poletti and Peter Horak.
\newblock {Description of ultrashort pulse propagation in multimode optical
  fibers}.
\newblock {\em J. Opt. Soc. Am. B}, 25(10):1645--1654, October 2008.

\bibitem{6242367}
A~Mafi.
\newblock {Pulse Propagation in a Short Nonlinear Graded-Index Multimode
  Optical Fiber}.
\newblock {\em Lightwave Technology, Journal of}, 30(17):2803--2811, 2012.

\bibitem{Shemirani2009}
Mahdieh~B. Shemirani, Wei Mao, Rahul~Alex Panicker, and Joseph~M. Kahn.
\newblock {Principal Modes in Graded-Index Multimode Fiber in Presence of
  Spatial- and Polarization-Mode Coupling}.
\newblock {\em Journal of Lightwave Technology}, 27(10):1248--1261, May 2009.

\bibitem{Shemirani2009a}
M.B. Shemirani and J.M. Kahn.
\newblock {Higher-Order Modal Dispersion in Graded-Index Multimode Fiber}.
\newblock {\em Journal of Lightwave Technology}, 27(23):5461--5468, December
  2009.

\bibitem{Shen:05}
Xiling Shen, Joseph~M Kahn, and Mark~A Horowitz.
\newblock {Compensation for multimode fiber dispersion by adaptive optics}.
\newblock {\em Opt. Lett.}, 30(22):2985--2987, November 2005.

\bibitem{Longhi2004}
Stefano Longhi and Davide Janner.
\newblock {Self-focusing and nonlinear periodic beams in parabolic index
  optical fibres}.
\newblock {\em Journal of Optics B: Quantum and Semiclassical Optics},
  6(5):S303--S308, May 2004.

\bibitem{PhysRevA.83.033838}
P~Aschieri, J~Garnier, C~Michel, V~Doya, and A~Picozzi.
\newblock {Condensation and thermalization of classsical optical waves in a
  waveguide}.
\newblock {\em Phys. Rev. A}, 83(3):33838, March 2011.

\bibitem{Barad1997}
Y.~Barad and Y.~Silberberg.
\newblock {Polarization Evolution and Polarization Instability of Solitons in a
  Birefringent Optical Fiber}.
\newblock {\em Physical Review Letters}, 78(17):3290--3293, April 1997.

\bibitem{Kang1996}
J.~Kang, G.~Stegeman, J.~Aitchison, and N.~Akhmediev.
\newblock {Observation of Manakov Spatial Solitons in AlGaAs Planar
  Waveguides}.
\newblock {\em Physical Review Letters}, 76(20):3699--3702, May 1996.

\bibitem{Chen1997}
Zhigang Chen, Mordechai Segev, Tamer~H. Coskun, Demetrios~N. Christodoulides,
  and Yuri~S. Kivshar.
\newblock {Coupled photorefractive spatial-soliton pairs}.
\newblock {\em Journal of the Optical Society of America B}, 14(11):3066,
  November 1997.

\bibitem{Pitaevskii2003}
L.~Pitaevskii and S.~Stringari.
\newblock {\em {Bose-Einstein Condensation (Physics)}}.
\newblock Oxford University Press, USA, 2003.

\bibitem{Agrawal1997}
Govind~P. Agrawal.
\newblock {\em {Fiber-Optic Communication Systems (Wiley Series in Microwave
  and Optical Engineering)}}.
\newblock Wiley-Interscience, 1997.

\bibitem{Berdague1982}
S.~Berdagu\'{e} and P.~Facq.
\newblock {Mode division multiplexing in optical fibers}.
\newblock {\em Applied Optics}, 21(11):1950, June 1982.

\bibitem{Stuart2000}
H.~R. Stuart.
\newblock {Dispersive Multiplexing in Multimode Optical Fiber}.
\newblock {\em Science}, 289(5477):281--283, July 2000.

\bibitem{Tarighat2007}
A.~Tarighat, R.C.J. Hsu, A.~Shah, A.H. Sayed, and B.~Jalali.
\newblock {Fundamentals and challenges of optical multiple-input
  multiple-output multimode fiber links [Topics in Optical Communications]}.
\newblock {\em IEEE Communications Magazine}, 45(5):57--63, May 2007.

\bibitem{Fermann1998}
Martin~E. Fermann.
\newblock {Single-mode excitation of multimode fibers with ultrashort pulses}.
\newblock {\em Optics Letters}, 23(1):52, January 1998.

\bibitem{Kivshar2003}
Y.~S. Kivshar and G.~P. Agrawal.
\newblock {\em {Optical solitons: From fibers to photonic crystals}}.
\newblock Academic Press, San Diego, 2003.

\bibitem{Mitschke1986}
F.~M. Mitschke and L.~F. Mollenauer.
\newblock {Discovery of the soliton self-frequency shift}.
\newblock {\em Optics Letters}, 11(10):659, October 1986.

\bibitem{Bespalov1978}
Bespalov, V.I., Betin, A.A., Pasmanik, and G.A.
\newblock {Reproduction of the pump wave in simulated scattering radiation}.
\newblock {\em Izvestiya Vysshikh Uchebnykh Zavedenii, Radiofizika},
  21(7):961--80, 1978.

\bibitem{Goldhar1982}
J.~Goldhar and J.~Murray.
\newblock {Intensity averaging and four-wave mixing in Raman amplifiers}.
\newblock {\em IEEE Journal of Quantum Electronics}, 18(3):399--409, March
  1982.

\bibitem{Chang1983}
R.~S.~F. Chang and N.~Djeu.
\newblock {Amplification of a diffraction-limited Stokes beam by a severely
  distorted pump}.
\newblock {\em Optics Letters}, 8(3):139, March 1983.

\bibitem{Essiambre2010}
Ren\'{e}-Jean Essiambre, Gerhard Kramer, Peter~J. Winzer, Gerard~J. Foschini,
  and Bernhard Goebel.
\newblock {Capacity Limits of Optical Fiber Networks}.
\newblock {\em Journal of Lightwave Technology}, 28(4):662--701, February 2010.

\end{thebibliography}

%

\section*{Acknowledgments}
Portions of this work were supported by the National Science Foundation (ECS-0901323 and PHYS-0653482).  The authors thank G. Agrawal and M. Segev for valuable discussions.

\section*{Author contributions}
W.H.R. performed experiments, simulations, and analysis.  F.W.W supervised the project.  The manuscript was prepared by W.H.R. and F.W.W.

%
%


%

\end{document}